\documentclass[fleqn,10pt]{wlscirep}

\usepackage{algorithm}
\usepackage{algpseudocode}

\algnewcommand{\algorithmicand}{\textbf{ and }}
\algnewcommand{\algorithmicor}{\textbf{ or }}
\algnewcommand{\algorithmicnot}{\textbf{not }}
\algnewcommand{\algorithmicset}{\textbf{set }}
\algnewcommand{\Or}{\algorithmicor}
\algnewcommand{\And}{\algorithmicand}
\algnewcommand{\Not}{\algorithmicnot}
\algnewcommand{\Set}{\algorithmicset}

\title{Virtual Testing of Experimental Continuation}

\author{Rainer M.J. Groh}
\author{Robin M. Neville}
\author{Alberto Pirrera}
\author{Mark Schenk}
\affil{Bristol Composites Institute (ACCIS), University of Bristol, BS8 1TR, Bristol, United Kingdom}

\begin{abstract}
We present a critical advance in experimental testing of nonlinear structures. Traditional quasi-static experimental methods control the displacement or force at one or more load-introduction points on a structure. This approach is unable to traverse limit points in the control parameter, as the immediate equilibrium beyond these points is statically unstable, causing the structure to snap to another equilibrium. As a result, unstable equilibria---observed numerically---are yet to be verified experimentally. Based on previous experimental work, and a virtual testing environment developed herein, we propose a new \textit{experimental continuation} method that can path-follow along unstable equilibria and traverse limit points. To support these developments, we provide insightful analogies between a fundamental building block of our technique---shape control---and analysis concepts such as the principle of virtual work and Galerkin's method. The proposed testing method will enable the validation of an emerging class of nonlinear structures that exploit instabilities for novel functionality.
\end{abstract}
\begin{document}

\flushbottom
\maketitle

\thispagestyle{empty}

\section*{Introduction}

In many physical and engineering disciplines, instabilities are viewed as a ``failure" mechanism. In engineering mechanics, an alternative perspective has developed over the last decade, whereby elastic instabilities are used to enable additional functionality~\cite{Hu2015,Reis2015}. For example, buckling has been used for applications as diverse as energy harvesting~\cite{Harne2013,Emam2015}, reversible shape-adaptation~\cite{Arena2017,Gomez2017}, surface texturing~\cite{Seffen2014}, actuation~\cite{Overvelde2015}, self-encapsulation~\cite{Shim2012}, auxetic materials~\cite{Bertoldi2010}, and energy dissipation~\cite{Hu2015c}. 

A fundamental building block that underpins these novel structures is the ubiquitous fold catastrophe, also known as the limit point or saddle-node bifurcation. In structural mechanics, the response of a structure is generally described by equilibrium curves of force \emph{vs} a chosen norm of the displacement field. When the structure reaches a \textit{limit point} in the loading parameter (force or displacement), the stability of the structure changes---a previously stable equilibrium becomes unstable, or \textit{vice versa}. Here, structural stability refers to the stability of the equilibrium with respect to small perturbations (Lyapunov stability).
This means that a structure that is loaded by a slowly evolving, yet monotonously increasing load (force or displacement control) will snap to another equilibrium upon reaching a limit point. Although force or displacement at the actuation point are controlled, the rest of the structure is free to move dynamically. While such snaps can be advantageous in terms of facilitating additional functionality (\emph{e.g.}\ shape adaptation), it leaves connecting unstable segments of the force-displacement equilibrium curve inaccessible to experimental testing.

Numerous numerical methods have been developed to analyse nonlinear structures. A common method, broadly classified under \emph{numerical continuation}, is based on the predictor-corrector scheme of Newton's method. To traverse limit points, force and displacement are decoupled by introducing an arc-length constraint equation~\cite{Riks1979}. In contrast, an experimental analogue to numerical continuation for quasi-static systems has remained elusive. Indeed, recent research has focused almost entirely on developing new analysis methods for nonlinear structures, but comparatively little progress has been made on new experimental methods. For example, current testing methods cannot follow unstable equilibria and cannot traverse limit points, meaning that several longstanding numerical benchmarks found in the literature are yet to be validated experimentally~\cite{Wardle2008}.

The aim of this paper is to outline an \emph{experimental continuation} method for quasi-static systems that overcomes some of the limitations of current testing methods. Analogous to the arc-length equation in numerical continuation, the fundamental---and enabling---building block that allows the decoupling of force and displacement in experimental continuation is \emph{shape control}~\cite{Neville2018}. Here, we explore the application of shape control to experimental continuation within a virtual testing environment.

\section*{Theory}
\begin{figure}[ht]
	\includegraphics{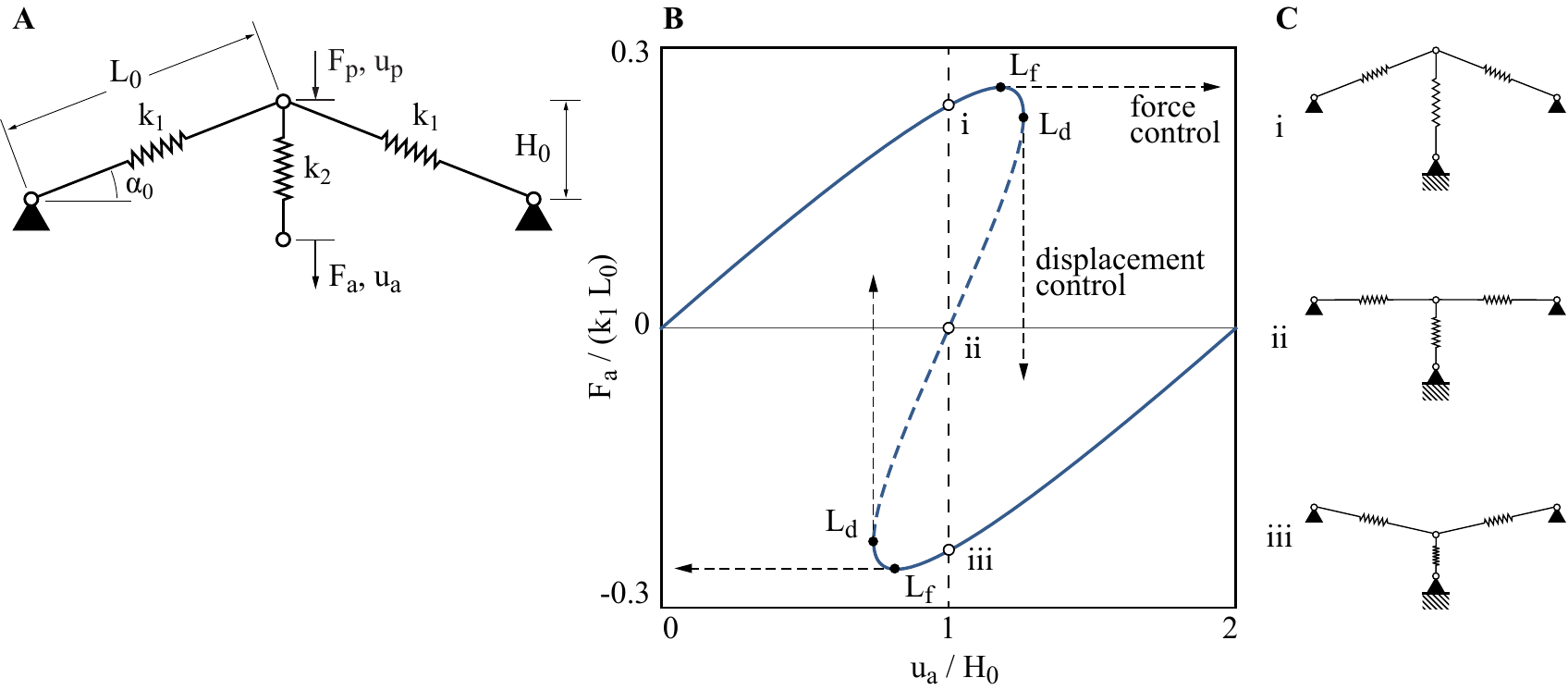}
    \centering
    \caption{(A) \emph{von Mises} truss geometry. The actuation point is at the bottom of the vertical spring; the probe point is at the apex of the truss. (B) The force-displacement curve for a \emph{von Mises} truss with $k_1/k_2 = 2$ and $\alpha_0 = 50^{\circ}$ has both displacement and force limit points. This causes the dashed section of the equilibrium curve to be unstable under either force or displacement control, and therefore inaccessible to quasi-static testing. The central section of the plot highlights three equilibria (points i, ii, and iii) that correspond to three forces ($F_\mathrm{a}$) associated with one displacement ($u_\mathrm{a}=H_0$). (C) Each equilibrium i, ii, and iii, and hence each force $F_\mathrm{a}$, shown in (B) is associated with a unique truss shape.}
    \label{fig:VM-truss}
\end{figure}

To illustrate the underlying concepts, we employ a simple structure that exhibits the salient features of nonlinear behaviour with limit points: the spring-loaded \emph{von Mises} truss (Figure~\ref{fig:VM-truss}). The \emph{von Mises} truss features an arch-like arrangement of two inclined linear springs, with a third spring suspended from the apex. For a load applied to the bottom of the vertical spring, the force-displacement ($F_\mathrm{a}$ \textit{vs} $u_\mathrm{a}$) response describes a general sigmoidal shape. The precise characteristics of this equilibrium curve are entirely described by the geometric arrangement ($\alpha_0$, $L_0$) and stiffness of the springs ($k_1/k_2$). For certain arrangements, the equilibrium curve features both force and displacement limit points (see Figure~\ref{fig:VM-truss}).

The section of the equilibrium curve bounded by the two displacement limit points (dashed segment in Figure~\ref{fig:VM-truss}B) is experimentally inaccessible using conventional techniques. This unstable segment of the equilibrium manifold acts as a repeller, whereas the two stable segments act as attractors. Hence, under displacement control, the apex snaps up- or downwards dynamically upon reaching the unstable segment. To path-follow along the unstable equilibrium segment and to traverse both limit points, a method for simultaneously controlling the force and displacement at the loading point is needed. The experimental challenge is that force and displacement are inherently linked through elasticity: an applied force results in a displacement, and an applied displacement induces a reaction force.

Crucially, within this unstable region, multiple equilibria (\emph{e.g.}\ points i, ii, and iii in Figure~\ref{fig:VM-truss}B) with different reaction force readings ($F_\mathrm{a}$) exist for each displacement of the actuation point ($u_\mathrm{a}$). What differentiates the different reaction force readings is the associated shape of the truss (see Figure~\ref{fig:VM-truss}C). This provides the key insight to experimentally decoupling force and displacement at the actuation point, namely introducing a third control variable: the overall geometric \emph{shape}. For an applied displacement, controlling the equilibrium shape determines the corresponding reaction force; conversely, for an applied force, the shape determines the displacement.

As shown in previous experimental work~\cite{Neville2018}, the shape of the structure can be controlled by introducing additional control points. The primary loading point(s) remain unchanged, but additional control point(s) are used to perturb the shape of the structure. In the case of the \emph{von Mises} truss, this is done by controlling the displacement of the apex ($u_\mathrm{p}$). Hence, the \emph{shape} of the structure is uniquely determined by the position of the apex \emph{and} the actuation point.
By controlling the deformation shape, the purpose of additional ``probe" point(s) is twofold. First, for unstable equilibria, the probes provide the stabilisation force against infinitesimal perturbations (stabilisation). Second, the probes can be used to select different equilibria that exist for a specific level of primary loading ($u_\mathrm{a}$ or $F_\mathrm{a}$). Each unique equilibrium state of the unprobed structure must correspond to zero reaction force at the probe points ($F_\mathrm{p} = 0$). When this is the case, as far as the structure is concerned, the probes ``do not exist". In effect, shape control decouples force and displacement at the primary actuation point, which permits the measurement of unstable equilibria and traversing of limit points.

This concept of obtaining a zero-force reading on the probes to pinpoint equilibria (stable and unstable) has two pertinent analogies in numerical methods: (i) the minimisation of virtual work in response to a probing virtual displacement; and (ii) the vanishing of the residual in Newton’s method. The principle of virtual work states that of all possible kinematically admissable (virtual) deformations, the one that minimises the total potential energy corresponds to the actual deformation. A powerful tool for solving the virtual work statement analytically/numerically is the Galerkin method, whereby kinematically admissable shape functions are assumed and the total residual over the domain is minimised. In precisely this fashion, the probes are used to impose a subset of the kinematically admissable displacements (the ones that can be controlled by the probes), and the residual is then minimised at specific points (\textit{i.e.}\ zero reaction force at the probe points). Similarly, most numerical frameworks used in structural mechanics---\emph{e.g.} finite difference or finite element methods---divide the computational domain into discretisation points (nodes). Some of these nodes are constrained from displacing (boundary conditions), others are loaded, and the rest are unloaded. As a reference load is applied, the unconstrained nodes displace, but in general, there is a difference between the induced internal nodal forces and the applied external nodal forces. Hence, the structure is not in equilibrium. In Newton's method, the structure is moved closer to an equilibrium state by applying the residual (the difference between internal and external nodal forces) as an additional force to \emph{all} nodes. As a result, previously unloaded nodes of the structure are now loaded, thereby controlling the overall shape of the structure beyond the primary actuation point. An equilibrium state is found when the residual falls beneath a predefined threshold. Again, the equivalent in experimental shape control is the vanishing reaction force at the probe points. 

Although the notion of using probe points to pinpoint unstable equilibria is gaining traction in the literature~\cite{Thompson2015b,Virot2017,Neville2018}, and has been used to path-follow along unstable equilibria~\cite{vanIderstein2019}, no algorithm yet exists that can traverse limit points. In the following, we delineate such an approach for simple structures and outline a roadmap to more complicated structures encountered in engineering practice.

\section*{Method}
We implement an experimental path-following algorithm in a virtual setting. The aim is to demonstrate what can be measured experimentally by modelling the experimental approach using finite element (FE) software. Using FE computations as the ``experimental measurement" allows us to explore, develop and validate different algorithms to be used in an experimental setting. Furthermore, it is possible to explore how measurement uncertainty will influence experimental readings. The model is implemented in the commercial FE software \textsc{Abaqus} using \textsc{Python} scripts, which have been supplied as part of the Supplementary Information (SI).

\begin{figure}[ht]
  \centering
  \includegraphics[width=\linewidth]{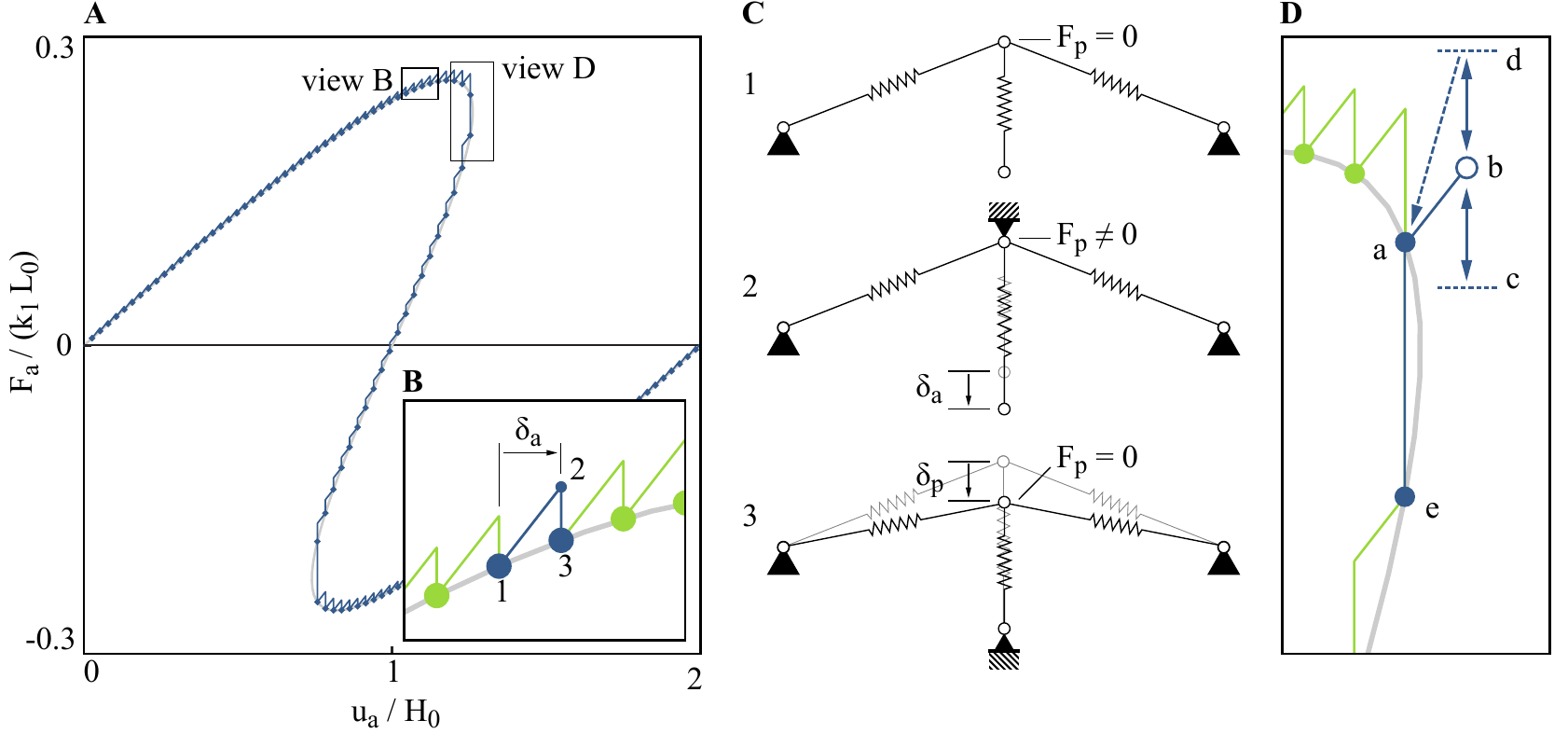}
  \caption{(A) The output of the experimental path-following algorithm (blue steps) superimposed on the equilibrium curve (grey). (B) Detail view of the stepping procedure described by Algorithm~\ref{alg:step-scan} in the SI. Starting at point 1 on the equilibrium curve, the probe is fixed and the actuation point is moved by increment $\delta_\mathrm{a}$, generating a non-zero probe force (point 2). The probe is then moved until zero probe force is found (point 3). The shapes associated with these points are shown conceptually in (C). (D) Detail view of the limit point tracing logic described by Algorithm~\ref{alg:limit-point-logic} in the SI. Starting at point $a$, we step by $\delta_\mathrm{a}$ to point $b$. The probe is then moved to search for zero probe force, but there are no solutions within bounds $c$ and $d$. The algorithm returns to the known equilibrium point $a$, and moves the probe until the next solution is found at point $e$.
  \label{fig:VM-continuation}}
\end{figure}

To implement a path-following algorithm, a combination of increasing the displacement at the actuation point(s), $u_\mathrm{a}$, and scanning for equilibria using the probe point(s), $u_\mathrm{p}$, is required. The first implementation of such an algorithm is here referred to as the \emph{step-scan} method. First, it involves a finite increment, or \emph{step}, $\delta_\mathrm{a}$ of the actuation point at constant probe displacement, $\delta_\mathrm{p} = 0$. This step (moving from 1 to 2 in Figure~\ref{fig:VM-continuation}) induces a nonzero reaction force at the probe point. Next, the probe point is moved by $\delta_\mathrm{p}$ until the probe reaction force reads zero (2 to 3 in Figure~\ref{fig:VM-continuation}), \textit{i.e.}\ the probe \textit{scans} for an equilibrium. This procedure is formally described by Algorithm~\ref{alg:step-scan} in the SI.

This basic \emph{step-scan} algorithm is used to progress along both statically stable and unstable equilibrium paths as shown graphically in Figure~\ref{fig:VM-continuation}B by the saw-tooth shaped segments, and schematically by the deformed shapes in Figure~\ref{fig:VM-continuation}C. As long as the actuation point increment, $\delta_\mathrm{a}$, is sufficiently small, the probe-scanning step (moving $\delta_\mathrm{p}$) will quickly encounter a zero reaction force, $F_\mathrm{p} = 0$, reading. However, this algorithm will break at a displacement limit point (see Figure~\ref{fig:VM-continuation}D). At the limit point, the control point increment, $\delta_\mathrm{a}$, takes the system into a region where the probe scanning step does not intersect an equilibrium solution. Hence, the probe reaction force will never reduce to zero ($F_\mathrm{p} \neq 0$). To overcome this, additional logic is required. Multiple options exist to traverse the limit point, yet, we propose the most basic of algorithms, whose simplicity should make it robust to experimental noise encountered in practice. 

Imagine incrementing the system from an equilibrium very close to the limit point as shown in Figure~\ref{fig:VM-continuation}D. To find a new equilibrium, the control system first probe-scans in one direction, and, upon failing to identify a zero reaction force reading within a preset bound, inverts to scan in the opposite direction. If a zero probe reaction force is also not found in this scanning direction, the control system returns the system back to its previously identified equilibrium, followed by a further probe scan in the original direction. This procedure is formally described by Algorithm~\ref{alg:limit-point-logic} in the SI. This improved \emph{step-scan} algorithm traverses both limit points observed in the force-displacement response of the \emph{von Mises} truss. Throughout this procedure, the controlling action of the probe prevents snapping, and the full mechanical response of the structure is made accessible experimentally.

\begin{figure}
\centering
	\includegraphics[width=\linewidth]{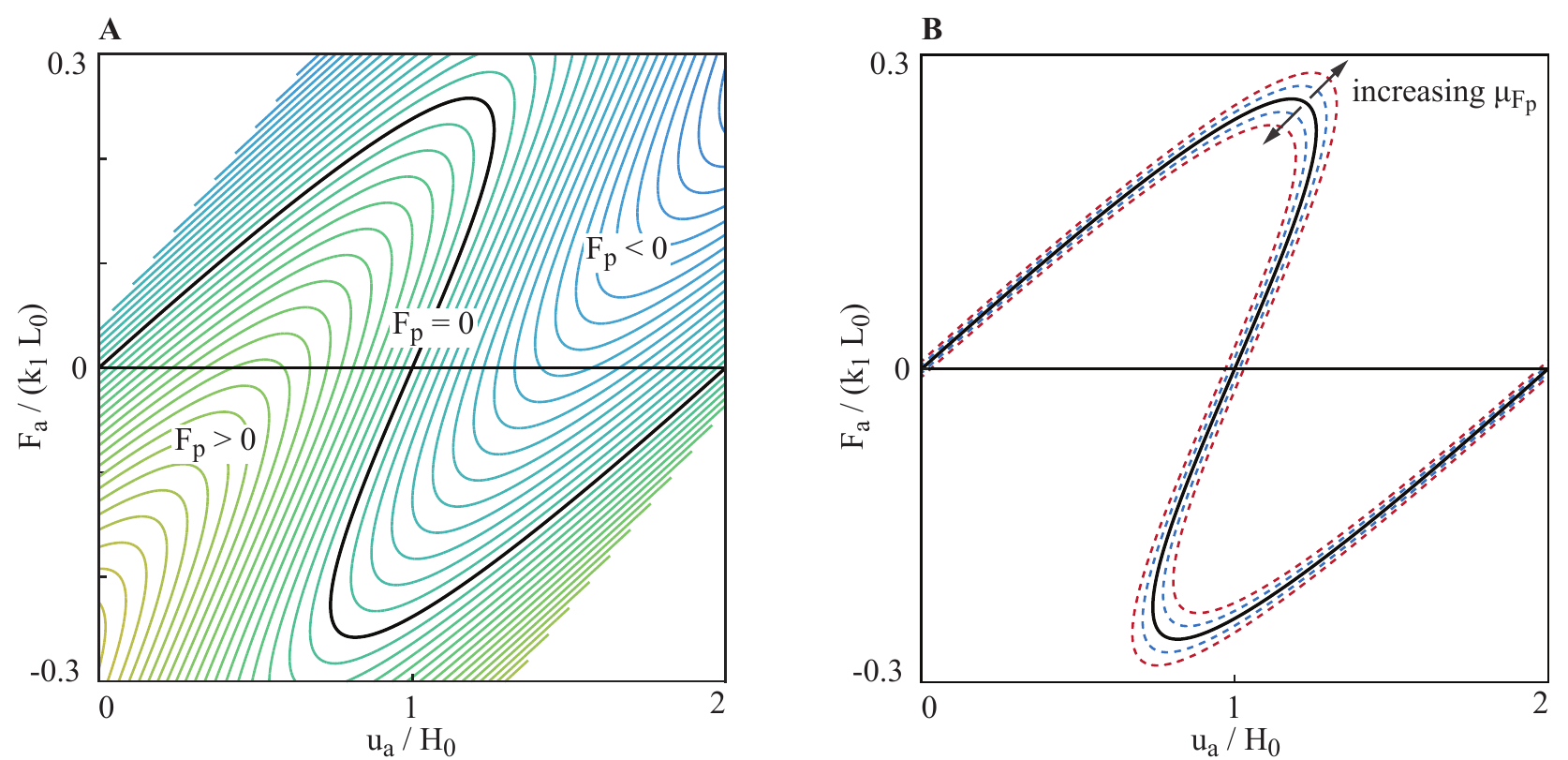}
    \caption{(A) By calculating the force at the actuation and probe point ($F_\mathrm{a}$ and $F_\mathrm{p}$) for every combination of actuation and probe point displacement ($u_\mathrm{a}$ and $u_\mathrm{p}$), a ``probe force manifold'' can be generated (the magnitude of $F_\mathrm{p}$ is indicated by the contour lines). The solid black equilibrium curve is found where the probe force is zero. (B) The manifold can be used to explore sensitivity to uncertainty in the probe force measurement. Shown are error bounds for $\mu_{F_\mathrm{p}}$ in increments of 5\% of the maximum measured actuation point force, $F^{\max}_\mathrm{a}$, around the known equilibrium solution.}
\label{fig:VMuncertainty}
\end{figure}

Although any physical implementation of the experiment will encounter additional factors in terms of imperfections, measurement error and noise, previous experiments by the present authors~\cite{Neville2018} have shown the simple \emph{step-scan} algorithm to work robustly. The additional logic of moving around a limit point builds upon this fundamental building block. 
Indeed, the sensitivity to measurement uncertainty can be quantified using the virtual testing environment. Taking the actuation point measurements (force and displacement) to be without error, a measurement uncertainty in the probe force ($\mu_{F_\mathrm{p}}$) results in an uncertainty in the measured equilibrium curve. Figure~\ref{fig:VMuncertainty} shows the true equilibrium curve ($F_\mathrm{p} = 0$) with increasing error bounds ($\mu_{F_\mathrm{p}} = 0.05 F^{\max}_\mathrm{a}$). The sensitivity to measurement uncertainty is greatest around the limit points, and smallest around the predominantly linear portions of the curve.

\section*{Discussion \& Outlook}
For the purpose of exposition, the discussion has thus far focused on the simplest structure that exhibits nonlinearity and limit points---the \emph{von Mises} truss. However, the proposed experimental continuation method is not restricted to this simplest of cases. For a general structure, one set of control points will still describe the physical loading imposed on the structure (actuation points), with another set of control points used to control the global deformation shape (probe points). The number of probe points required is determined by the minimum number needed to control a given unstable equilibrium shape. 

\begin{figure}
	\centering
	\includegraphics[width=\linewidth]{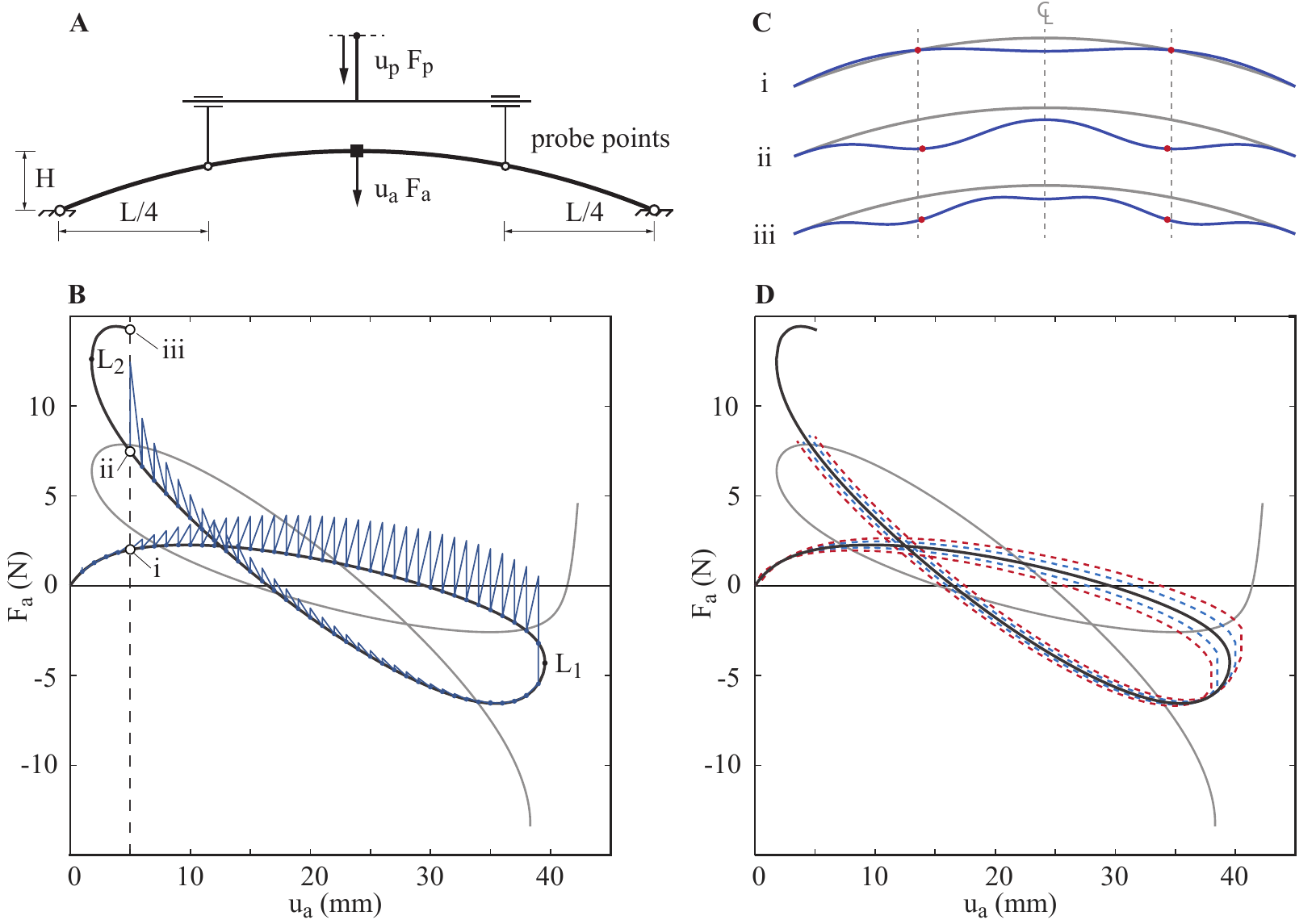}
    \caption{(A) a shallow symmetric arch, with probe locations at $L/4$ from both supports; for further details see reference~\cite{Neville2018}. (B) Result of path-following algorithm with a 1\,mm step size, which traverses limit point L$_1$ but breaks down near the second limit point L$_2$, due to lack of control authority over the deformation shape of the structure. (C) The deformed shapes of the arch for equilibrium positions i, ii and iii for $u_\mathrm{a} = 5$\,mm, with probe locations highlighted. The current number of probe points and locations cannot control shape iii, and hence additional probes are required to traverse beyond the second limit point L$_2$. (D) Probe force measurement uncertainty bounds (dashed lines) in uncertainty increments of $\mu_{F_\mathrm{p}} = \pm 0.25$\,N around the known equilibrium solution (solid line).}
\label{fig:arch}
\end{figure}

For instance, the scanning algorithm has been applied to a symmetric shallow arch using two probe points and one actuation point~\cite{Neville2018}. Figure~\ref{fig:arch}A shows the shallow arch with a midspan actuation point and two probe points at quarter and three-quarter span. In this case, the probes provide the stabilisation force to prevent snapping, and additionally enforce symmetry. Figure~\ref{fig:arch}B shows the result of the path-following algorithm; the solid line is the equilibrium curve computed from a numerical arc-length solver. The virtual experimental continuation framework has no difficulty in traversing the first limit point (L$_1$), but cannot go beyond the second limit point (L$_2$). This second limit point separates deformation mode shapes with five and seven half-waves (see Figure~\ref{fig:arch}C). The experimental continuation setup does not possess sufficient deformation fidelity, \emph{i.e.}\ number and location of probes, to control the higher-order deformation mode shape, and this results in a loss of ``control authority''. This was of no concern for the \textit{von Mises} truss, as the probe provided full control over the deformation shape. The \textit{a priori} selection of the number and location of probes is an important factor in successful experimental continuation, as these will determine which equilibria can be identified. The virtual testing environment offers the ability to explore these factors before commencing an experimental programme. Lastly, Figure~\ref{fig:arch}D shows the probe force measurement uncertainty bounds in increments of $\pm 0.25$\,N around the known equilibrium solution. The measurement sensitivity is a function of the structural mechanics, as well as the choice of probe locations.

The notion of ``control authority" suggests that with increasing structural complexity (\emph{e.g.}\ from arch to shell), the number of probe points to control the structure increases correspondingly. A practical experimental continuation setup under these circumstances requires a more sophisticated control algorithm. Ideally, one that moves all control points in tandem based on the non-zero reaction force readings at all probe points, \emph{i.e.}\ an experimental analog to numerical continuation. As described previously, Newton's method formulates a corrective loading step based on the tangential stiffness matrix and residual forces at discretisation points. By analogy, an experimental tangential stiffness matrix may be established by perturbing a single probe by a small amount and recording the change in reaction force at all probes. By repeating this procedure for all probes sequentially, a finite-difference tangential stiffness matrix can be assembled. Consequently, the residual force readings at probe points and the tangential stiffness matrix can be used to compute a corrector displacement for all probe points that concertedly moves them towards the unprobed equilibrium. In this manner, the testing method can not only be scaled up to larger structures, but the algorithm also departs from the rudimentary \emph{step-scan} approach to a full-fledged experimental continuation method.

\section*{Conclusion}
This paper presents a significant advance in experimental methods for nonlinear structures. The proposed \textit{experimental continuation} method can path-follow along nonlinear equilibrium curves and traverse limit points in a structure's force-displacement response. The fundamental building block (shape control) of the proposed method has been demonstrated experimentally previously~\cite{Neville2018}. Here, we have explored the testing algorithm virtually via a finite element implementation of the experimental method, and demonstrated its capabilities. The virtual testing environment is used to gain insight into the measurement uncertainty of the unstable equilibrium paths. Further, we outline a means of extending the experimental continuation method to more complex structures, by exploiting elegant analogies with \emph{numerical} continuation methods. Fundamentally, there exists a direct mapping of numerical quantities, such as the tangential stiffness matrix and the residual vector, to the experimental domain. In the near future, we expect that a full-fledged experimental continuation approach will significantly enhance engineers' and scientists' capabilities to test and validate buckling-driven multifunctional structures.

\section*{Acknowledgements}
An initial implementation of the \textsc{ABAQUS} analyses was written by Toby Cornwall, as part of an undergraduate summer internship funded by the Bristol Composites Institute (ACCIS). RMJG is supported by the Royal Academy of Engineering under the Research Fellowship scheme [RF\textbackslash201718\textbackslash17178]. AP is funded by the Engineering and Physical Sciences Research Council (EPSRC) under their Research Fellowship scheme [EP/M013170/1]. RN is supported by the EPSRC under grant number [EP/N509619/1]. The support of all funders is gratefully acknowledged.

\section*{Data Statement}
The \textsc{Python} scripts to run the virtual testing environment for the \emph{von Mises} truss and the shallow arch can be downloaded from the data repository of the University of Bristol via the URL: \textcolor{blue}{https://doi.org/10.5523/bris.2lfdqgcpnpdd82673rbnegw34w}

\clearpage

\appendix

\section{Supplementary Information}

\begin{algorithm}
\caption{Basic \emph{Step-Scan} Path-following} \label{alg:step-scan}
\begin{algorithmic}[1]
\Loop
\State $u_\mathrm{a} = u_\mathrm{a} + \delta_\mathrm{a} \And u_\mathrm{p} = \text{constant}$
\Repeat
\State $u_\mathrm{p} = u_\mathrm{p} + \delta_\mathrm{p} \And u_\mathrm{a} = \text{constant}$
\Until{$F_\mathrm{p} = 0$}
\EndLoop
\end{algorithmic}
\end{algorithm}

\begin{algorithm} 
\caption{Improved \emph{Step-Scan} Path-following} \label{alg:limit-point-logic}
\begin{algorithmic}[1]
\Require $\left(u_\mathrm{a},u_\mathrm{p}\right) _\mathrm{eq}$
\Loop
\State $u_\mathrm{a} = u_\mathrm{a} + \delta_\mathrm{a} \And u_\mathrm{p} = \text{constant}$
\While{$u_\mathrm{p} < \|u_\mathrm{p}^{\mathrm{bound}}\|$}
\Repeat
\State $u_\mathrm{p} = u_\mathrm{p} + \delta_\mathrm{p} \And u_\mathrm{a} = \text{constant}$
\Until{$F_\mathrm{p} = 0$}
\EndWhile
\If{$\Not F_\mathrm{p} = 0$}
\While{$u_\mathrm{p} > -\|u_\mathrm{p}^{\mathrm{bound}}\|$}
\Repeat
\State $u_\mathrm{p} = u_\mathrm{p} - \delta_\mathrm{p} \And u_\mathrm{a} = \text{constant}$
\Until{$F_\mathrm{p} = 0$}
\EndWhile
\EndIf
\If{$\Not F_\mathrm{p} = 0$}
\State \Set $\left(u_\mathrm{a},u_\mathrm{p}\right) = \left(u_\mathrm{a},u_\mathrm{p}\right) _\mathrm{eq} \And \delta_\mathrm{a} = -\delta_\mathrm{a}$
\State $u_\mathrm{a} = u_\mathrm{a} + 0.01\delta_\mathrm{a} \And u_\mathrm{p} = \text{constant}$
\Repeat
\State $u_\mathrm{p} = u_\mathrm{p} + \delta_\mathrm{p} \And u_\mathrm{a} = \text{constant}$
\Until{$F_\mathrm{p} = 0$}
\EndIf
\EndLoop
\end{algorithmic}
\end{algorithm}

\newpage

\bibliographystyle{unsrt}
\bibliography{bibliography}
\end{document}